\documentclass{aa} %

\makeatletter
\let\numberlines@hook\relax
\makeatother

\usepackage[varg]{txfonts}

\DeclareRobustCommand{\VAN}[3]{#2}
\let\VANthebibliography\thebibliography
\def\thebibliography{\DeclareRobustCommand{\VAN}[3]{##3}\VANthebibliography}

\usepackage{graphicx}	
\usepackage{amsmath}	
\usepackage[version=4]{mhchem}     

\usepackage{mathtools}
\DeclarePairedDelimiter{\diagfences}{(}{)}
\newcommand{\diag}{\operatorname{diag}\diagfences}

\usepackage{hyperref}
\hypersetup{colorlinks=true,linkcolor=blue,citecolor=blue,filecolor=blue,urlcolor=blue, breaklinks=True}

\usepackage{booktabs}   

\usepackage[referable,para,flushleft]{threeparttablex} 

\newcommand{\kepler}{\textit{Kepler}\xspace}

\newcommand{\msol}{\ensuremath{\mathrm{M}_\odot}\xspace}

\newcommand{\Zsol}{\ensuremath{\mathrm{Z}_\odot}\xspace}

\makeatletter
\newcommand\thefontsize[1]{{#1 The current font size is: \f@size pt\par}}
\makeatother

\begin{document}

\title{Anomalously low-mass core-He-burning star in NGC 6819 as a post-common-envelope phase product}

\author{Massimiliano Matteuzzi\inst{\ref{inst1},\ref{inst2}}\thanks{E-mail: \href{mailto:massimilia.matteuzz2@unibo.it}{massimilia.matteuzz2@unibo.it}}
\and
David Hendriks\inst{\ref{inst3}}
\and
Robert G. Izzard\inst{\ref{inst3}}
\and
Andrea Miglio\inst{\ref{inst1},\ref{inst2}}
\and
Karsten Brogaard\inst{\ref{inst1},\ref{inst2}}
\and
Josefina Montalb\'an\inst{\ref{inst1}}
\and
Marco Tailo\inst{\ref{inst1}}
\and
Alessandro Mazzi\inst{\ref{inst1}}
}

\institute{Department of Physics \& Astronomy "Augusto Righi", University of Bologna, via Gobetti 93/2, 40129 Bologna, Italy\label{inst1}
\and
INAF-Astrophysics and Space Science Observatory of Bologna, via Gobetti 93/3, 40129 Bologna, Italy\label{inst2}
\and
Department of Physics, University of Surrey, Guildford, Surrey GU2 7XH, UK\label{inst3}
}

\abstract{
Precise masses of red-giant stars enable a robust inference of their ages, but there are cases where these age estimates are highly precise yet very inaccurate. Examples are core-helium-burning (CHeB) stars that have lost more mass than predicted by standard single-star evolutionary models. Members of star clusters in the \kepler database represent a unique opportunity to identify such stars, because they combine exquisite asteroseismic constraints with independent age information (members of a star cluster share similar age and chemical composition). In this work we focus on the single, metal-rich ($Z \approx \Zsol$), Li-rich, low-mass, CHeB star KIC4937011, which is a member of the open cluster NGC 6819 (turn-off mass of $\approx 1.6$ \msol, i.e. age of $\approx 2.4$ Gyr). This star has $\approx 1$ \msol less mass than expected for its age and metallicity, which could be explained by binary interactions or mass-loss along the red-giant branch (RGB). To infer formation scenarios for this object, we perform a Bayesian analysis by combining the binary stellar evolutionary framework {\scshape binary\_c v2.2.3} with the dynamic nested sampling approach contained in the {\scshape dynesty v2.1.1} package. We find that this star is likely the result of a common-envelope evolution (CEE) phase during the RGB stage of the primary star in which the low-mass ($<0.71$ \msol) main sequence companion does not survive. The mass of the primary star at the zero-age main sequence is in the range $[1.46, 1.71]$ \msol, with a log-orbital period in the range $[0.06, 2.4] \, \log_{10}(\mathrm{days})$. During the CEE phase $\approx 1 \,  \msol$ of material is ejected from the system, and the final star reaches the CHeB stage after helium flashes as if it were a single star of mass $\approx 0.7 \, \msol$, which is what we observe today. Although the proposed scenario is consistent with photometric and spectroscopic observations, a quantitative comparison with detailed stellar evolution calculations is needed to quantify the systematic skewness of radius, luminosity, and effective temperature distributions towards higher values than observations.
}



\titlerunning{Anomalously low-mass core-He-burning star in NGC 6819 as a post-common-envelope phase product}
\authorrunning{Matteuzzi et al.}
\maketitle

\section{Introduction}
\label{sec:intro}
The advent of the space mission \kepler has made it possible to study in great detail the oscillation spectra of tens of thousands of red-giant stars, and thus, to constrain their internal structures, properties and evolutionary phases. These asteroseismic constraints, coupled with information on photospheric chemical abundances and temperature, have also given us the ability to precisely measure their radii and masses \citep[][]{DeRidder2009,Hekker2011,Huber2011,Miglio2013,Stello2013,Mosser2014,Yu2018,Garcia2019,Kallinger2019}.

Precise masses of red-giant stars enable a robust inference of their ages \citep[][]{Anders2016,Casagrande2016,Pinsonneault2018,SilvaAguirre2018,Miglio2021,Montalban2021}, but there are cases in the field and in open star clusters where these estimates are significantly different than the age predicted by standard single-star evolutionary models. For example, some members of open clusters have more mass than the observed average mass for their evolutionary phase \citep{Brogaard2016,Handberg2017,Brogaard2018,Brogaard2021}. This means that such stars appear younger than they actually are, thus, they must have experienced mass transfer or merger events in the past \citep[][]{Izzard2018}.
Despite potential insights from observations that could assist in distinguishing between different formation scenarios \citep[][]{Brogaard2018}, the evolutionary history of many observed systems with masses that deviate from their expected mass remains uncertain. The limited support from other stars in age determination makes it challenging to identify non-standard evolutionary paths for field stars. An exception to this is the thick disc, which has a well-defined turn-off mass that allows for the tracking of the evolutionary history of these stars \citep[][]{Chiappini2015,Martig2015,Izzard2018,Grisoni2024}.

While there are observed systems with a significant excess of mass with respect to the  average mass in the cluster for their evolutionary phase, the opposite is true as well. There are red-giant members of open clusters with less mass than the observed average mass for their evolutionary phase \citep{Handberg2017,Brogaard2021}. Considering their mass and metallicity, they have lost more mass than expected, most likely via interaction with a companion star \citep[][]{Li2022,Bobrick2024} or because of mass loss along the red-giant branch (RGB). There are observations of such undermassive stars also in the field \citep[][]{Miglio2021,Li2022,Matteuzzi2023}. Some of them are low-mass, core-helium-burning (CHeB) stars located in the colour-magnitude diagram between the RR Lyrae and the red clump (RC), thus are red horizontal branch (rHB) stars. By modelling their structure and pulsation spectra, \citet[][]{Matteuzzi2023} found that these low-mass objects have a helium-core mass of $\approx 0.5$ \msol and a hydrogen-rich envelope of $\approx 0.1 - 0.2$ \msol, that is they are stars with a less massive envelope than other stars in the same evolutionary phase and with similar metallicity, but slightly more massive than the RR Lyrae stars with similar metallicity. This means that such undermassive stars are partially stripped, probably as a result of a past binary interaction. Investigating plausible formation scenarios for these stars is critical to better constrain their actual ages and to potentially provide another piece of the puzzle in the sequence between RC stars, metal-rich RR Lyrae and subdwarf B (sdB) stars, or other stripped stars.

In binary stars, mass transfer is possible by direct Roche-lobe overflow (RLOF) or by wind mass loss \citep[see][for a review]{DeMarco2017}. When mass is transferred from red-giant stars and low-mass main sequence (MS) stars on a dynamical time-scale, the companion may be engulfed and a common-envelope evolution begins \citep[CEE; e.g.][]{Paczynski1976,Ivanova2013b,Ropke2023}. This can happen when mass-transfer is unstable, i.e. the transfer of mass by the donor leads to an increase in mass-transfer rate. Whether the mass-transfer becomes unstable is usually determined based on a critical mass-ratio, $q_{\rm crit}$, between the primary mass and the companion mass and when the system exceeds this mass ratio \citep[$q_1> q_{\rm crit}$,][]{Hurley2002} it will undergo unstable mass-transfer followed by a CEE. At this stage drag forces transfer part of the orbital energy to the common envelope (CE), shrinking the orbit and ejecting at least part of the CE \citep[][]{Shima1985,Kim2010,MacLeod2015,Ohlmann2016,Chamandy2019,Reichardt2019,Sand2020}. The consequence of this CE phase is that stars either merge, or end up much loser than before this phase. The easiest way to model the CEE is using the $\alpha$-formalism \citep[][]{vandenHeuvel1976,Webbink1984,Livio1988,deKool1990,Han1994,Dewi2000,Xu2010a,Xu2010b,Ivanova2011a,Wang2016}, also called the energy formalism. This formalism can be described by the following equation,
\begin{equation}
\label{eq:CEE}
\centering
E_{\rm bind}  = \alpha_{\rm ce} \Delta E_{\rm orb},
\end{equation}
where
\begin{equation}
\label{eq:CEE2}
\centering
E_{\rm bind} = - \frac{G m_1 m_{\rm 1,env}} {\lambda_{\rm ce} R_1}
\end{equation}
and 
\begin{equation}
\label{eq:CEE3}
\centering
\Delta E_{\rm orb} = - \frac{Gm_{\rm 1,core}m_2}{2a_{\rm f}}  + \frac{G m_1m_2}{2a_{\rm i}}
\end{equation}
are the binding energy of the envelope of the primary star (the donor star) and the difference in orbital energy after and before the CEE, respectively. $E_{\rm bind}$ depends on the mass ($m_1$), the envelope mass ($m_{\rm 1,env}$), the Roche-lobe radius ($R_1$) of the primary star, and it also contains a numerical factor $\lambda_{\rm ce}$ to characterise the central concentration of the envelope. $\Delta E_{\rm orb}$ depends also on the core mass of the primary star ($m_{\rm 1,core} = m_1 - m_{\rm 1,env}$), the mass of the companion star (the accretor star), and the orbital separation before ($a_{\rm i}$) and after ($a_{\rm f}$) the CEE. Finally, $\alpha_{\rm ce}$ indicates the fraction of orbital energy converted into energy used to eject the envelope, thus it is the efficiency of the CE ejection. Unfortunately, there are no direct observations of CEE events with which we can constrain $\lambda_{\rm ce}$ and $\alpha_{\rm ce}$. Nonetheless, observations of post-CEE systems and 3D hydrodynamical simulations suggest that the value of $\alpha_{\rm ce}$ is not universal and depends on many factors such as donor mass, mass ratio and evolutionary stage \citep[][]{Taam2000,Podsiadlowski2003,Politano2004,DeMarco2011,Davis2012,Iaconi2019,Belloni2024}. Moreover, many works \citep[][]{Han1994,Dewi2000,Dewi2001,Podsiadlowski2003,Webbink2008,Xu2010a,Xu2010b,Wong2014} suggest that $\lambda_{\rm ce}$ varies as the star evolves and significantly deviates from a constant value. However, from an analysis of equations \ref{eq:CEE}, \ref{eq:CEE2} and \ref{eq:CEE3}, it is evident that the properties of the post-CEE system do not change when the product $\alpha_{\rm ce} \cdot \lambda_{\rm ce}$ is held constant. Despite our limited understanding of the variables $\lambda_{\rm ce}$ and $\alpha_{\rm ce}$, we can effectively constrain the post-CEE phase by using $\alpha_{\rm ce} \cdot \lambda_{\rm ce}$. Conversely, a better knowledge of the post-CEE phase gives us a constraint on the possible $\alpha_{\rm ce} \cdot \lambda_{\rm ce}$ values.

Members of star clusters observed by the \kepler space telescope represent a unique opportunity to constrain formation channels of undermassive stars, because they combine exquisite asteroseismic constraints with age information since members of a star cluster share similar age and chemical composition. In this paper we focus on the red-giant star KIC4937011, a member of the Galactic star cluster NGC 6819 which has a turn-off mass of $\approx 1.6$ \msol \citep[i.e. an age of $\approx 2.4$ Gyr,][]{Burkhead1971,Lindoff1972,Auner1974,Rosvick1998,Kalirai2001,Basu2011,Yang2013,Jeffries2013,Bedin2015,Brewer2016}. Based on the scaling relation between the mean large frequency separation and the frequency of maximum power, \citet[][]{Handberg2017} determined that KIC4937011 has a mass of $0.71 \pm 0.08$ \msol. This value is consistent, within the errors, with the mass calculated using the scaling relation involving the luminosity and the frequency of maximum power \citep[][]{Matteuzzi2023}. Additionally, analysis of the asymptotic period spacing of the dipole modes indicates that KIC4937011 is currently in the RC phase \citep[][]{Matteuzzi2023}. Based on spectroscopy \citep[][]{Anthony-Twarog2013,Carlberg2015,Lee-Brown2015}, we know that it is a single star with near-solar metallicity, a high lithium and oxygen content, and a rotational velocity of $8.3 \pm 0.3$ km/s (i.e. a higher value than the other red-giant stars in this cluster). This star has $\approx 1$ \msol less mass than the average mass of RC stars in NGC 6819 \citep[i.e. $1.64$ \msol,][]{Handberg2017}, which could be explained by binary interactions.

The paper is organised as follows. We briefly describe the observational properties of KIC4937011 in Section \ref{sec:obs}. In Section \ref{sec:MCsim} we describe the theoretical framework that permits us to predict the formation channel, and how we simulate binary interactions. In Section \ref{sec:results} we present our results, namely the most credible formation channels for this star, which is the consequence of a common-envelope evolution phase in which the companion does not survive. Section \ref{sec:conc} concludes.

\section{Observational data}
\label{sec:obs}
\begin{figure*}
\centering
\includegraphics{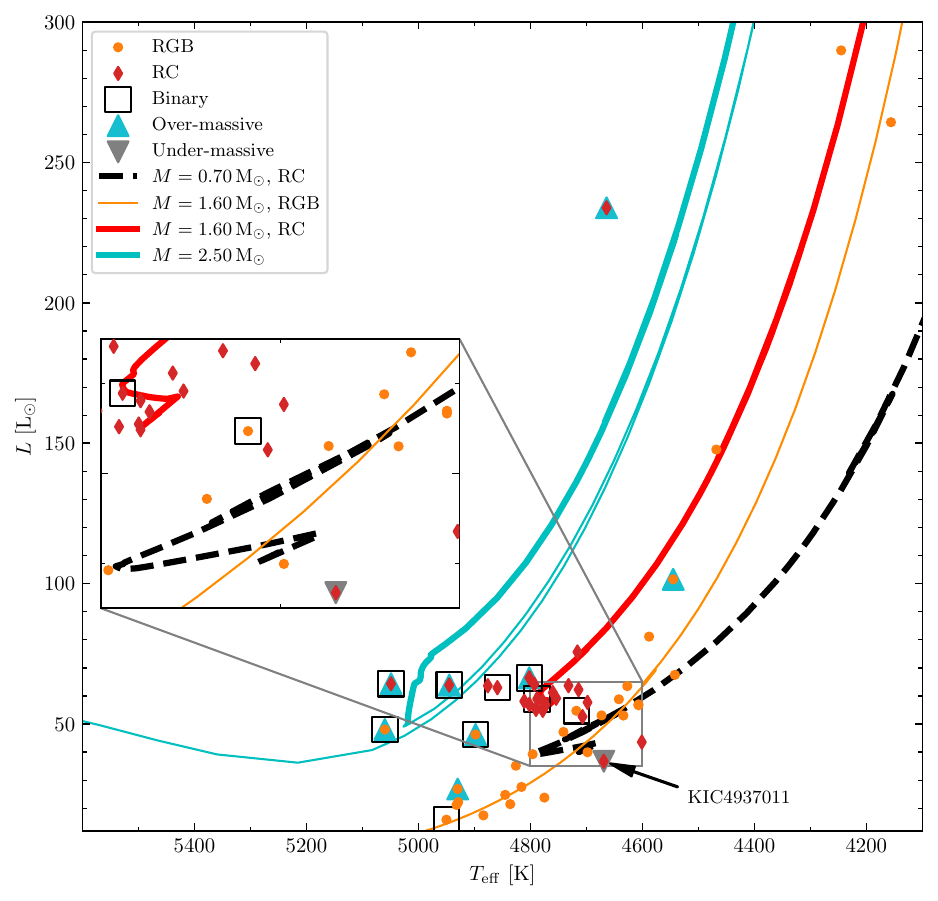} 
\caption{HRD of NGC 6819 showing RC and RGB member stars. Apart from standard single stars, some are members of binary systems, some are overmassive stars and one (KIC4937011) is undermassive. The solid lines are evolutionary tracks from the RGB phase until the first thermal pulse in the asymptotic giant branch (AGB) phase for a 1.60 \msol and a 2.50 \msol star with solar metallicity, while the dotted line represents a 0.70 \msol star with solar metallicity from the beginning of the CHeB stage until the first thermal pulse in the AGB phase. KIC4937011 is less bright and hot than other member stars at the same evolutionary stage, and it would be compatible with a 1.60 \msol RGB star if we had no information about its evolutionary state and mass. For a full description of the models used in the figure see \citet{Handberg2017}.}
\label{fig:HRD} 
\end{figure*}

In Section \ref{sec:intro} we state that KIC4937011 is a RC member of the open cluster NGC 6819 (HRD in Figure \ref{fig:HRD}, observational properties in Table \ref{tab:fullsample}), and this is suggested by its radial velocity \citep[][]{Hole2009,Anthony-Twarog2013,Carlberg2015}, CMD position \citep[][]{Anthony-Twarog2013}, and proper-motion \citep{GaiaCollaboration2016,GaiaCollaboration2023,Babusiaux2023}. This allows for an independent estimation of the age of KIC4937011 regardless of its mass. In this paper we use the age estimate $2.38 \pm 0.05 \pm 0.22 $ Gyr\footnote{These uncertainties are estimates of the random and systematic effects (due to model physics differences and metal content), respectively.} based on eclipsing binaries of \citet{Brewer2016}, but with an error of 0.27 Gyr\footnote{We do not add the two uncertainties in quadrature, because the direct addition is an upper limit of the true uncertainty (Schwarz inequality), which does not depend on any possible correlation between the two.} (Section \ref{sec:modeling}).
The metallicity of the cluster is similar to solar \citep[][]{Rosvick1998,Anthony-Twarog2014,Lee-Brown2015,Slumstrup2019}, however other works suggest super-solar metallicities \citep[$ \mathrm{[Fe/H]} = 0.09 \pm 0.03$,][]{Bragaglia2001}. This is in line with the metallicity of KIC4937011, which is approximately solar ($\mathrm{[Fe/H]} = 0.04$, \citeauthor{Carlberg2015} \citeyear{Carlberg2015}; $\mathrm{[Fe/H]} = -0.05$, \citeauthor{Lee-Brown2015} \citeyear{Lee-Brown2015}). In this paper we use a metallicity $Z = 0.02$ for KIC4937011 (Section \ref{sec:modeling}). The observed \ce{^{12}C/^{14}N} of KIC4937011 is consistent with its other more massive ($\mathrm{i.e.} \approx 1.64$ \msol) counterparts in the RC phase, although the lower the mass, the higher \ce{^{12}C/^{14}N} should be in a star that evolves in isolation \citep[$\ce{^{12}C/^{14}N} > 3$ for such a low-mass, metal-rich star,][]{Salaris2015}. Past mass transfer events may provide an explanation for the peculiar \ce{^{12}C/^{14}N} observed in the envelope of KIC4937011. It is possible that the envelope originates from a star that during its first dredge-up was more massive than the current KIC4937011 \citep[][]{Hekker2019,Tayar2023}.

\citet{Handberg2017} derive masses in NGC 6819 in the RGB and RC phases using asteroseismic data from \kepler, obtaining average masses of $1.61 \pm 0.02$ \msol and $1.64 \pm 0.02$ \msol, respectively. This means that the integrated mass loss in the RGB phase in this cluster is compatible with a Reimers' mass loss law \citep{Reimers1975,Reimers1978} that has an efficiency $\eta_{\rm RGB} = 0.1$. However, KIC4937011 has almost $1$ \msol less mass than its other counterparts in the RC phase. As a result, an efficiency $\eta_{\rm RGB} > 1$ is required in order to explain the isolated evolution of this star. Such a high efficiency is highly improbable, as it far exceeds the average value measured in NGC 6819. Therefore, it is unlikely that mass loss in the RGB phase through winds alone can account for the mass discrepancy observed in KIC4937011. Indeed, \citet{Carlberg2015} proposed an interaction between a $\approx 1.7$ \msol red-giant star and a brown dwarf of mass $45 \, \mathrm{M_{Jup}}$ to explain the enrichment in lithium $[\mathrm{A(Li)} = 12 + \log \left( N_\mathrm{Li}/N_\mathrm{H} \right) = +2.3$ dex, \citeauthor{Anthony-Twarog2013} \citeyear{Anthony-Twarog2013}] and the loss of almost $1$ \msol. However, alternative explanations exist for the observed Li-enrichment in red-giant stars. Some observations suggest that R-stars with similar chemical composition and lithium content may have interacted with a companion star instead of a brown dwarf \citep[][]{Zamora2006}. Another possibility is that lithium-rich RC stars could have been created through the merging of a helium white dwarf (HeWD) star and a RGB star \citep[][]{Zhang2020}. Several mechanisms have been proposed to account for the Li-enrichment in red-giant stars, including planet or brown dwarf engulfment \citep[][]{Ashwell2005,AguileraGomez2016a,AguileraGomez2016b}, accretion from an asymptotic giant branch (AGB) star or a nova \citep[][]{Jose1998}, and the \citet{Cameron1971} mechanism with some extra-mixing \citep[][]{Denissenkov2004,Guandalini2009,Denissenkov2010,AguileraGomez2023}, such as the HeWD-RGB merger mentioned earlier. Therefore, the lithium content alone is not a reliable indicator to discriminate between formation channels. However, considering all available pieces of information, we need to incorporate binary interactions into evolutionary models of KIC4937011.

\citet{Carlberg2015} perform multiple checks of whether the spectroscopic analysis is affected by either a companion to the Li-rich star or an unrelated background object. They find that the radial velocity is constant within $0.1 - 0.2$ km/s in nearly one month of observations. Even when they consider 20 years of observations with other surveys they do not obtain variations in its radial velocity. They also search for a secondary spectrum, but they do not identify any significant secondary peaks. Furthermore, there is no indication of infrared excess, which is observed in other Li-rich red-giant stars \citep[][]{Rebull2015,Mallick2022}. Even using the currently available Gaia-DR3 astrometry data \citep[non-single star processing in][]{Halbwachs2023} there is no evidence for companion stars. We also test the non-single star hypothesis using the {\scshape fidelity\_v2} table \citep[][]{Rybizki2022} and the {\scshape RUWE} value\footnote{The renormalised unit weight error ({\scshape RUWE}) is the square root of the normalised chi-square of the astrometric fit to the along-scan observation. It is expected to be $1.0$ for well-behaved solutions of single stars \citep{GaiaCollaboration2016,GaiaCollaboration2023}.}. Together, all this information suggests that KIC4937011 has no companion.

\section{Bayesian inference of formation scenarios}
\label{sec:MCsim}
In this section we describe the Bayesian approach we adopt to infer the most probable formation scenario for KIC4937011. This is done using an evolutionary code for binary stars (Section \ref{sec:binary_c}) coupled with a Monte Carlo (MC) method (Section \ref{sec:modeling}).

\subsection{Evolutionary code for binary stars}
\label{sec:binary_c}
The software {\scshape binary\_c v2.2.3}\footnote{\url{https://gitlab.com/binary_c/binary_c/-/tree/releases/2.2.3}} \citep{Izzard2004,Izzard2006,Izzard2009,Izzard2018,Izzard2023} makes synthetic populations of single, binary and multiple stars. It is based upon the Binary Star Evolution ({\scshape BSE}) code \citep{Hurley2002}, which uses analytic fits to rapidly follow the properties of a system as a function of time \citep[][]{Hurley2000}. In addition, {\scshape binary\_c v2.2.3} rapidly calculates nucleosynthetic yields from such synthetic populations, adopting first, second, third dredge-up and thermally-pulsing asymptotic giant branch (TPAGB) prescriptions from \citet{Izzard2006}, which are based on \citet[][]{Karakas2002}, and the supernova yields from massive stars from \citet{Woosley1995, Chieffi2004}. These prescriptions are very useful for globular cluster and Galactic chemical evolution simulations \citep[][]{Izzard2018,Yates2024}. The physics implemented in the code \citep[][for a review about relevant physical processes in binary systems]{DeMarco2017} is mainly described in the above cited papers. The code allows for the incorporation of alternative models beyond those utilised in the BSE code, such as different RLOF \citep{Claeys2014}, wind Roche-lobe overflow \citep[WRLOF;][]{Abate2013,Abate2015}, accretion and thermohaline mixing \citep{Stancliffe2007,Izzard2018}, supernovae \citep{Boubert2017a,Boubert2017b}, tides \citep{Siess2013}, rejuvenation \citep{deMink2013, Schneider2014}, stellar rotation \citep{deMink2013}, stellar lifetimes \citep{Schneider2014}, CEE \citep{Wang2016}, and circumbinary discs \citep{Izzard2023}. This algorithm operates about $10^7$ times faster than full evolution and nucleosynthesis calculations, which makes it very useful for a Bayesian approach to parameter estimation.

In this paper we adopt a Python interface to {\scshape binary\_c v2.2.3}, {\scshape binary\_c-python v0.9.6}\footnote{\url{https://gitlab.com/binary_c/binary_c-python/-/tree/releases/0.9.6/2.2.3}} \citep{Hendriks2023}. We mainly adopt the binary-physics prescriptions of the BSE code, but the model of the properties of the AGB comes from \citet{Karakas2002}, the RLOF modelling onto a white dwarf from \citet{Claeys2014}, and the critical mass ratios $q_{\rm crit}$ from Table \ref{tab:qcrit}.
\begin{table}
\centering
\caption{Critical mass ratio $q_{\rm crit}$ for stable RLOF for different types of donor stars, in the case of a non-degenerate and a degenerate accretor. The meaning of each acronym is in \citet{Hurley2002}.}
\label{tab:qcrit}
\addtolength{\tabcolsep}{-0.2em}
\begin{tabular*}{\linewidth}{@{}lll@{}}
\toprule
Donor  & Non-deg accretor & Deg accretor\\ \midrule
MS with $M \le 0.718$ \msol  &  0.6944 &  1.0 \\ \midrule
MS with $M > 0.718$ \msol &  1.6 & 1.0 \\ \midrule
HG, HeHG & 4.0 & 4.7619 \\ \midrule 
RGB, EAGB, TPAGB & \citet{Hurley2002}  & 1.15 \\ \midrule
CHeB, HeMS  &  3.0 & 3.0 \\ \midrule
HeGB  &  0.78125 & 1.15 \\ \midrule
WD, NS, BH  &  3.0 & 0.625 \\ \bottomrule
\end{tabular*}
\end{table}
We fix to 0.5 the fraction of the recombination energy in the CE that participates in the ejection of the envelope (i.e. $\lambda_{\rm ionisation} = 0.5$ in {\scshape binary\_c v2.2.3}), and we consider $\alpha_{\rm ce} \cdot \lambda_{\rm ce}$ as a single free parameter of the model. We force a first dredge-up in Hertzsprung gap (HG) and RGB stars that undergo a CEE phase, but not for MS stars, because the dynamical mixing effects owing to the spiral-in process are assumed to completely mix the envelope in red-giant stars \citep{Izzard2006}. Furthermore, we do not include mass loss enhanced by rotation, tides and He flashes; we ignore thermohaline mixing, WRLOF and the lithium abundance change over time. 

\subsection{Monte Carlo simulations}
\label{sec:modeling}
To constrain binary systems that best explain the current state of KIC4937011, we need to efficiently estimate the posterior of a set of parameters for a given model obtained with {\scshape binary\_c v2.2.3} and {\scshape binary\_c-python v0.9.6}. We use the dynamic nested sampling approach contained in the {\scshape dynesty v2.1.1}\footnote{\url{https://github.com/joshspeagle/dynesty/tree/v2.1.1}} package \citep{Speagle2020}.

Initial conditions of all MC simulations are a binary system formed by zero-age main sequence (ZAMS) stars with circular orbits, $Z = 0.02$ and $\eta_{\rm RGB} = 0.1$, where we choose the metallicity and mass loss to be compatible with observations (Section \ref{sec:obs}). Our main results do not change when we allow $Z$ and/or $\eta_{\rm RGB}$ to vary freely within the observational uncertainties. The main difference is that the density distributions become broader and less predictive. We do not consider initially eccentric binaries, because the evolution of close binary populations is almost independent of the initial eccentricity \citep[][]{Hurley2002}.

We use uniform priors for $\alpha_{\rm ce} \cdot \lambda_{\rm ce}$ (Sections \ref{sec:intro} and \ref{sec:binary_c}), the logarithm\footnote{In this paper we use the notation $\log (x) \equiv \log_{10} (x) $.} of the initial period, $\log P_0$, and the initial mass ratio, $q_{\rm ZAMS} = M_{\rm 2, ZAMS} / M_{\rm 1, ZAMS}$. For the initial mass of the primary star, $M_{\rm 1, ZAMS}$, we employ the probability density distribution of \citet{Chabrier2003}\footnote{We check that uniform priors for $M_{\rm 1, ZAMS}$ lead to the same results. However such new priors require more CPU time.}. We calculate the likelihood function given the current evolutionary phase (CHeB), mass ($M_{\rm 1, CHeB} = 0.71 \pm 0.08$ \msol) and age ($t_{\rm age} = 2.38 \pm 0.27$ Gyr) of the star (Sections \ref{sec:intro}, \ref{sec:obs}). For more details on the likelihood used and the intervals chosen for our priors we refer the reader to Appendix \ref{app:Likelihood} and Table \ref{tab:priorsrange}.
\begin{table}
\centering
\caption{Medians and credible intervals (CI) of the density distributions explained in Section \ref{sec:modeling} and \ref{sec:results_fullsample}. In the last column the observed values for KIC4937011 with the corresponding references. In Section \ref{sec:chemistry} we discuss where such narrow CI in chemistry come from.}
\label{tab:fullsample}
\begin{tabular}{@{}p{0.2\columnwidth}p{0.1\columnwidth}p{0.2\columnwidth}p{0.2\columnwidth}p{0.1\columnwidth}@{}}
\toprule
Parameter  & Median & 68\% CI & 99.7\% CI & Obs\\ \midrule
$t_{\rm age}$ [Gyr]  &  2.40 & [2.18, 2.62] & [1.43, 3.18] & $2.38 \pm 0.27$\tablefootmark{a}  \\ \midrule
$M_{\rm 1, CHeB}$ [$\mathrm{M_\odot}$]  &  0.74 & [0.65, 0.83] & [0.50, 0.96] & $0.71 \pm 0.08$\tablefootmark{b} \\ \midrule
$\alpha_{\rm ce} \cdot \lambda_{\rm ce}$ & 14 & [6, 23] & [0.5, 47] & ... \\ \midrule
$\log \left(P_0 / \mathrm{days}\right)$ & 0.58  & [0.36, 1.06] & [0.08, 2.2] & ...\\ \midrule
$q_{\rm ZAMS}$  &  0.29 & [0.16, 0.37] & [0.05, 0.55] &  ... \\ \midrule
$M_{\rm 1, ZAMS}$ [$\mathrm{M_\odot}$]  &  1.69 & [1.59, 1.87] & [1.48, 2.62] & ...  \\ \midrule
$M_{\rm 2, ZAMS}$ [$\mathrm{M_\odot}$]  &  0.50 & [0.29, 0.63] & [0.081, 0.88] & ...  \\ \midrule
$M_{\rm 1, He-core}$  [$\mathrm{M_\odot}$] &  0.48 & [0.38, 0.53] & [0.32, 0.60] & ...\\ \midrule
$R_\mathrm{1, CHeB}$ [$\mathrm{R_\odot}$]  &  20 & [9.6, 28] & [0.2, 50] & $9.3 \pm 0.5$\tablefootmark{b} \\ \midrule
$L_\mathrm{1, CHeB}$ [$\mathrm{L_\odot}$]  &  148 & [65, 191] & [40, 376] & $37 \pm 4$\tablefootmark{b} \\ \midrule
$T_\mathrm{eff, 1, CHeB}$ [K]  &  4270 & [4074, 5347] & [3588, 27556] &  $4710 \pm 50$\tablefootmark{b} \\ \midrule
\ce{^{12}C/^{13}C}  &  90 & [32, 90] & [14, 90] & $25 \pm 5$\tablefootmark{d} \\ \midrule
\ce{^{12}C/^{14}N}  &  3.2 & [1.4, 3.2] & [0.62, 3.2] &  $1.4 \pm 0.2$\tablefootmark{d} \\ \midrule
\ce{^{14}N/^{16}O}  &  0.13 & [0.13, 0.23] & [0.13, 0.34] & $0.17 \pm 0.04$\tablefootmark{d} \\ \bottomrule
\end{tabular}
\tablebib{
\tablefoottext{a}{\citet[][]{Brewer2016};}
\tablefoottext{b}{\citet[][]{Handberg2017};}
\tablefoottext{b}{\citet[][]{Matteuzzi2023};}
\tablefoottext{d}{\citet[][]{Carlberg2015}}
}
\end{table}
\begin{figure}
\centering
\includegraphics{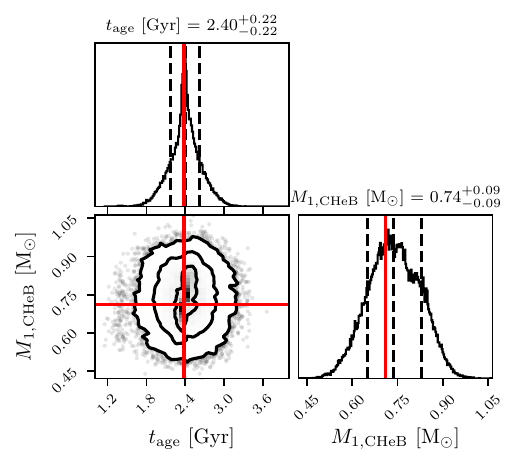} 
\caption{Corner plot showing the age and primary mass density distributions at the CHeB stage for our full sample described in Section \ref{sec:modeling}. The contours are referred to 1, 2, and 3-$\sigma$ credible regions, respectively. In red the observed values for KIC4937011 (see also Table \ref{tab:fullsample}). We see that our modelling correctly fits, within errors, the observed current age and mass of KIC4937011. As explained in the Appendix \ref{app:Likelihood}, we take as the reference time for the CHeB of each MC simulation the age that gives the highest likelihood.}
\label{fig:cornerdyn_fit}
\end{figure}

In Fig. \ref{fig:cornerdyn_fit} the corner plot showing the age and primary-mass density distributions of our MC simulations at the CHeB stage (see also Table \ref{tab:fullsample}), compared with the observed values for KIC4937011 (red lines). Our modelling correctly fits, within errors, the observed current age and mass of KIC4937011.

\section{Results}
\label{sec:results}
In this section we present significant results derived from our Monte Carlo simulations (Section \ref{sec:results_fullsample}), and we discuss the most credible formation channel (Section \ref{sec:results_subsample}).

\subsection{Formation channels constrained by age and mass observations}
\label{sec:results_fullsample}
\begin{figure*}
\centering
\includegraphics{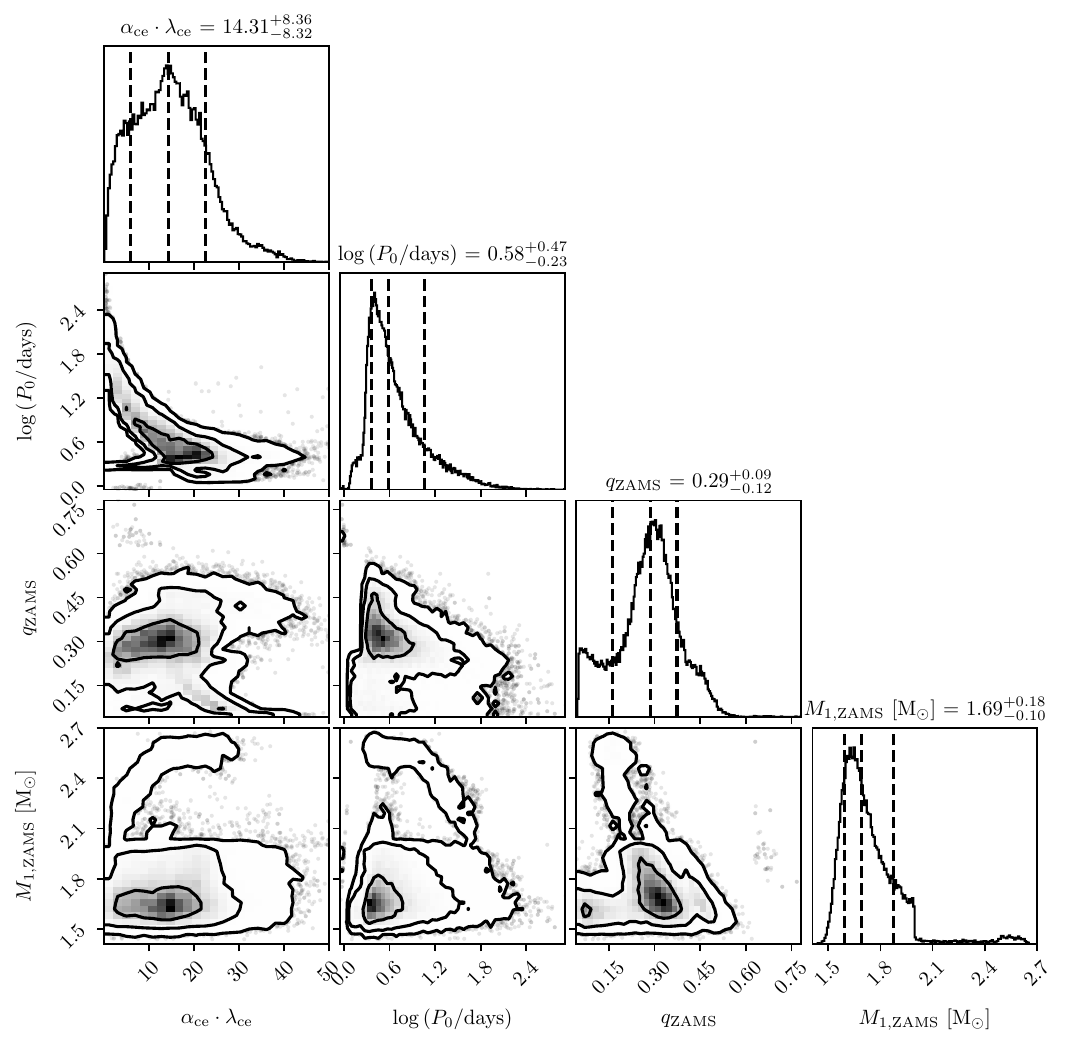} 
\caption{Corner plot showing the posterior density distributions of our MC simulations described in Section \ref{sec:modeling}. The contours are referred to 1, 2, and 3-$\sigma$, respectively.}
\label{fig:cornerdyn}
\end{figure*}

Figure~\ref{fig:cornerdyn} shows the estimated posterior density distributions of Monte Carlo simulations discussed in Section \ref{sec:modeling} and presented in Table \ref{tab:fullsample}.
A strong anti-correlation is observed between the product $\alpha_{\rm ce} \cdot \lambda_{\rm ce} $ and $\log \, P_0$, with a Pearson correlation coefficient of approximately $-0.69$. There are also weaker anti-correlations between $\log P_0$ and the other parameters. Equation \ref{eq:CEE} indicates that as the period increases, so does the orbital radius, while the surface gravity of the CE decreases, requiring a lower $\alpha_{\rm ce}$ for mass ejection. This suggests that certain areas of the $\log \, P_0 - \alpha_{\rm ce} \cdot \lambda_{\rm ce}$ plane can be excluded (Figure \ref{fig:cornerdyn} and Table \ref{tab:fullsample}), but we are unable to provide stricter constraints on the individual values of $\log \, P_0$ and $\alpha_{\rm ce} \cdot \lambda_{\rm ce} $. Furthermore, $\alpha_{\rm ce} \cdot \lambda_{\rm ce} $ here tends to higher values than suggested by other systems (Section \ref{sec:intro}). A plausible physical interpretation for this phenomenon relates to recombination energy. Specifically, high $\alpha_{\rm ce} \cdot \lambda_{\rm ce} $ values can be obtained when $\alpha_{\rm ce}$ is constrained to values below one, and $\lambda_{\rm ce}$ is calculated according to the \citet{Wang2016} prescriptions, utilising a high $\lambda_{\rm ionisation}$ value. Nonetheless, we cannot rule out the influence of additional energy sources, including dust formation or nuclear burning, that occur during the CEE phase. They may also contribute to the observed high $\alpha_{\rm ce} \cdot \lambda_{\rm ce} $ values.

Our estimated $M_{\rm 1, ZAMS}$ (Figure \ref{fig:cornerdyn} and Table \ref{tab:fullsample}) is slightly higher than, but still consistent with, the observed average mass of RGB stars in NGC6819 \citep[$1.61 \pm 0.02$ \msol,][]{Handberg2017}. This implies that a RGB star with a mass of 1.87 \msol (i.e. the upper limit of our 68\% credible interval for $M_{\rm 1, ZAMS}$) would begin the RGB phase about 0.82 Gyr\footnote{This age difference is calculated using {\scshape binary\_c v2.2.3} at $Z = 0.02$ and $\eta_{\rm RGB} = 0.1$.} earlier than the current RGB stars in NGC6819. Therefore, any binary interaction and evolution happened between a primary RGB star and a companion should last less than $\approx 1$ Gyr to be consistent with the observations. Moreover, the $q_{\rm ZAMS}$ posterior distribution lower limit in the 99.7\% credible interval is very close to the lower limit of our prior distribution (Table \ref{tab:priorsrange}), and probably limits in the prior below $0.08$ \msol are necessary for the companion mass.

\subsubsection{Common-envelope phases create distinct pathways}
\label{sec:ncom}
In this section we summarise the results concerning the number of CEE phases up to the CHeB stage of the primary star for the full sample. Every MC simulation considered (i.e. 65225 simulations) had one CEE phase, with a negligible percentage (0.006\%, i.e. 4 out of 65225 simulations) having two. This implies that our modelling predicts KIC4937011 to be a post-common-envelope phase product. The majority of the binary systems (i.e. 99.92\%) evolves without mass transfers until the primary star is in the subgiant or in the RGB phase and the companion in the MS phase. At this point an unstable mass transfer begins, leading to a CEE phase, and to the shrinkage of the orbits (Section \ref{sec:intro}). The loss of orbital energy is the reason why the probability of undergoing another CEE phase decreases significantly. Finally, in all our sample of MC simulations the companion star merges with the primary star, leaving as the final product a single CHeB star. This is in line with the observations (Section \ref{sec:obs}).

\subsubsection{Primary star physical properties}
\label{sec:core_distrib1}
\begin{figure*}
\centering
\includegraphics
{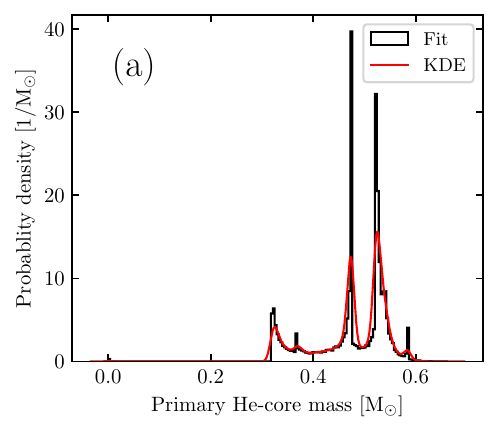}
\includegraphics{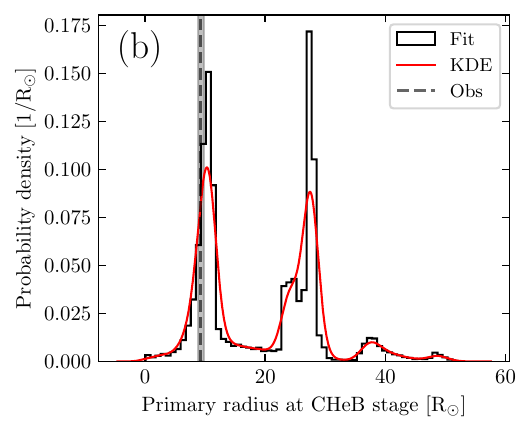}
\includegraphics{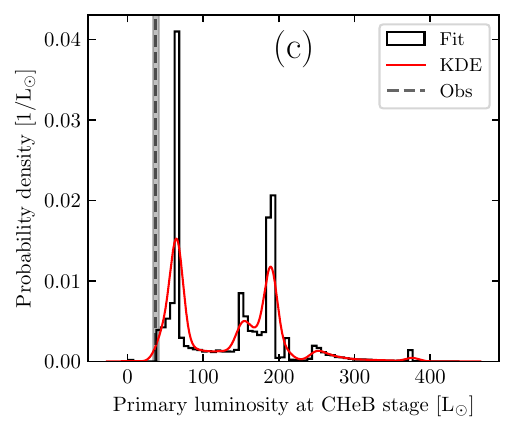}
\includegraphics{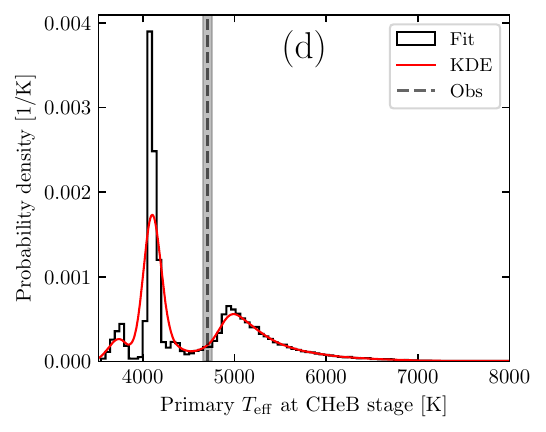}
\caption{Helium-core mass (panel a), radius (panel b), luminosity (panel c) and effective temperature (panel d) posterior density distributions of the primary star during the CHeB stage for the full sample described in Section \ref{sec:modeling}. We show the histogram (black lines) and the kernel density estimate with a Gaussian kernel (KDE, red lines). The dashed line and the grey area represent the KIC4937011 observations and their 1-$\sigma$ errors (Table \ref{tab:fullsample}). Effective temperatures above 8000 K have been omitted from the figure for illustrative purposes only. As explained in the Appendix \ref{app:Likelihood}, we take as the reference time for the CHeB of each MC simulation the age that gives the highest likelihood.}
\label{fig:core_distrib1}
\end{figure*}
In Fig.~\ref{fig:core_distrib1} and Table \ref{tab:fullsample} we present the helium-core mass, radius, luminosity, and effective temperature posterior density distributions of the primary star at the CHeB stage estimated from the MC simulations. In our simulations the primary stars have helium-core masses in the 99.7\% credible interval $[0.32, 0.60]$ \msol. Moreover, from the panel (a) we see that there are three main peaks: at around $0.32$ \msol, $0.47$ \msol and $0.52$ \msol. The first helium-core mass peak is expected for a secondary clump star, while the second for a RC star \citep[][]{Girardi2016}. However, helium-core masses above $\approx 0.50$ \msol are expected during the early asymptotic-giant branch (EAGB) stage and not during the CHeB stage. Such high helium-core masses lead to much higher radii ($> 15\, R_\odot$) and luminosities ($>100 \, L_\odot$) than expected of CHeB stars of mass $\approx 0.7$ \msol, and they also lead to effective temperatures well below 4500 K. Moreover, such high luminosities would not be consistent with the observed mixed modes behaviour in KIC4937011. It is worth noting that {\scshape binary\_c v2.2.3} is based on the \citet{Pols1998} evolutionary models, which predict higher radii, luminosities, and effective temperatures for such low-mass CHeB stars than more recent evolutionary models \citep[][]{Girardi2016}. These systematic effects must be taken into account when comparing with observations (Section \ref{sec:new_posteriors}).

\subsubsection{Dichotomy in the chemical space}
\label{sec:chemistry}
\begin{figure}
\centering
\includegraphics{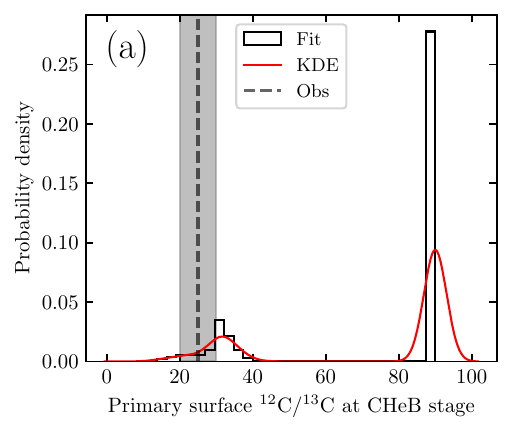} 
\includegraphics{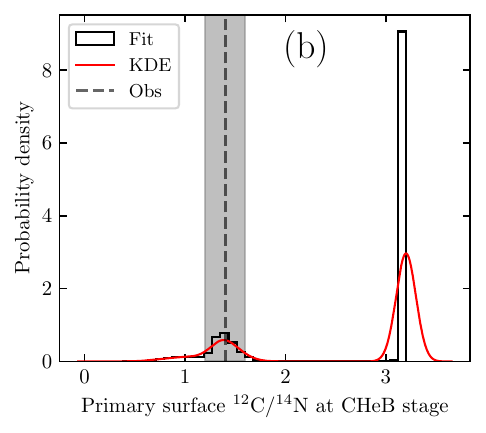} 
\includegraphics{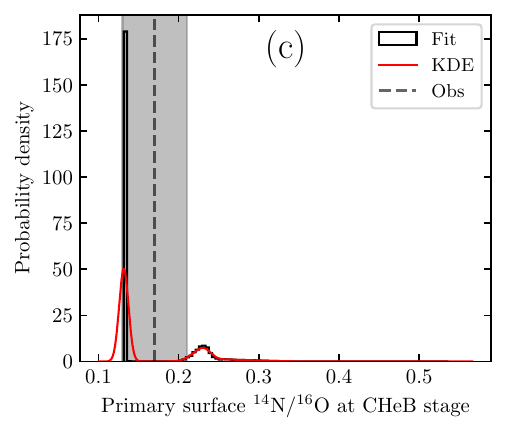} 
\caption{Surface \ce{^{12}C/^{13}C} (panel a), \ce{^{12}C/^{14}N} (panel b) and \ce{^{14}N/^{16}O} (panel c) posterior density distributions of the primary star at the CHeB stage for our full sample described in Section \ref{sec:modeling}. In particular, we have the histogram (black lines) and the kernel density estimation with a Gaussian kernel (KDE, red lines). The dashed line and the grey area represent the observed values of KIC4937011 and their 1-$\sigma$ errors (also Table \ref{tab:fullsample}). As explained in the Appendix \ref{app:Likelihood}, we take as the reference time for the CHeB of each MC simulation the age that gives the highest likelihood.}
\label{fig:chemistry}
\end{figure}
In Fig. \ref{fig:chemistry} and Table \ref{tab:fullsample} the \ce{^{12}C/^{13}C}, \ce{^{12}C/^{14}N} and \ce{^{14}N/^{16}O} density distributions of the primary star at the CHeB stage estimated from the MC simulations. A dichotomy is clearly visible in the chemical space, since simulations with the highest \ce{^{12}C/^{13}C} also have the highest \ce{^{12}C/^{14}N} and the lowest \ce{^{14}N/^{16}O}. These two peaks are the consequence of two different paths a binary system can follow after the CEE phase. In this section we discuss the channel that produces the highest \ce{^{12}C/^{14}N} peak, while in section \ref{sec:formation_channel} we discuss the channel that produces the lowest \ce{^{12}C/^{14}N} peak.

To explain the peak at $\ce{^{12}C/^{13}C} \approx 90 $, we need a star that contains material in the surface that has not been contaminated by the product of the hydrogen-burning core. Indeed, such a value is not typically observed in CHeB stars, because the first dredge-up has already taken place. As explained in Section \ref{sec:ncom}, the CEE phase begins when the binary systems are formed by a red-giant primary star and a MS companion. A first dredge-up is forced in HG and RGB stars that undergo a CEE phase, but not for MS stars (Section \ref{sec:binary_c}). In addition, MS stars are assumed to have negligible contamination from companion material, as all such material is ejected from the system soon after the CEE phase. These are the reasons why in our modelling we obtain that 69.51\% of the full sample of MC simulations (i.e. 45339 out of 65225 simulations) after the CEE phase are close binaries formed by a HeWD star and a MS star with $M < 0.88$ \msol\footnote{Formally this is not the same as the 99.7\% credible interval of $M_{\rm 2, ZAMS}$ in Table \ref{tab:fullsample}, but we checked whether the companion star has not changed much since the ZAMS.} that has not been contaminated by the primary star.
Such close-binary systems start a stable RLOF from the MS star onto the HeWD star, which eventually formes a single low-mass RGB-like star that later becomes a low-mass CHeB star with a high $\ce{^{12}C/^{13}C}$. This formation channel explains such a wide distribution of helium-core mass, radius, luminosity, and effective temperature, despite a narrow distribution in chemistry; the envelope comes from the MS star, thus, it depends more on the chosen initial composition than on internal mixing, atomic diffusion or overshooting. We want to highlight that the initial chemical composition of all these MC simulations is taken from \citet{Grevesse1989}, thus it is fixed for a fixed metallicity (this explains such narrow bins).

The properties of this formation channel are in good agreement with the literature, because binary systems formed by a HeWD star and a low-mass MS star are thought to be progenitors of low-mass RGB stars and of hot subdwarfs depending on the initial properties of the binary system \citep[][]{Hurley2002,Shen2009,Clausen2011,Pyrzas2012,Nelemans2016,Zhang2017,Rui2024}.

\subsection{Analysis of a more observationally-motivated subsample}
\label{sec:results_subsample}
\begin{figure}
\centering
\includegraphics[width=\columnwidth]{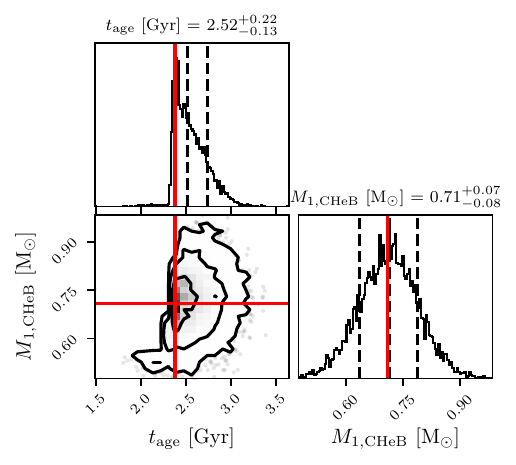}
\caption{Corner plot showing the age and primary mass density distributions at the CHeB stage for the subsample described in Section \ref{sec:results_subsample}. The contours are referred to 1, 2, and 3-$\sigma$, respectively. In red the observed values for KIC4937011 (Table \ref{tab:subsample}). As explained in the Appendix \ref{app:Likelihood}, we take as the reference time for the CHeB of each MC simulation the age that gives the highest likelihood.}
\label{fig:cornerdyn_teff1_sub}
\end{figure}
\begin{table}
\caption{The same as Table \ref{tab:fullsample}, but for the subsample described in Section \ref{sec:results_subsample}.}
\label{tab:subsample}
\addtolength{\tabcolsep}{-0.55em}
\begin{tabular}{@{}p{0.25\columnwidth}p{0.15\columnwidth}p{0.25\columnwidth}p{0.2\columnwidth}p{0.14\columnwidth}@{}}
\toprule
Parameter  & Median & 68\% CI & 99.7\% CI & Obs\\ \midrule
$t_{\rm age}$ [Gyr]  &  2.52 & [2.39, 2.74] & [1.99, 3.22] & $2.38 \pm 0.27$\tablefootmark{a}  \\ \midrule
$M_{\rm 1, CHeB}$ [$\mathrm{M_\odot}$]  &  0.71 & [0.64, 0.79] & [0.50, 0.94] & $0.71 \pm 0.08$\tablefootmark{b} \\ \midrule
$\alpha_{\rm ce} \cdot \lambda_{\rm ce}$ & 14 & [6, 24] & [0.4, 43] & ...\\ \midrule
$\log \left(P_0 / \mathrm{days}\right)$& 0.54 & [0.26, 1.06] & [0.06, 2.4] & ...  \\ \midrule 
$q_{\rm ZAMS}$  &  0.15 & [0.08, 0.30] & [0.05, 0.46] &...  \\ \midrule
$M_{\rm 1, ZAMS}$ [$\mathrm{M_\odot}$]  &  1.58 & [1.54, 1.61] & [1.46, 1.71] & ...  \\ \midrule
$M_{\rm 2, ZAMS}$ [$\mathrm{M_\odot}$]  &  0.24 & [0.12, 0.48] & [0.080, 0.71] & ...  \\ \midrule
$M_{\rm 1, He-core}$  [$\mathrm{M_\odot}$] &  0.4743 & [0.4735, 0.4783] & [0.4677, 0.5059] & ... \\ \midrule
$\rm R_{1, CHeB}$ [$\mathrm{R_\odot}$]  &  9.8 & [7.7, 10.9] & [0.38, 13.3] & $9.3 \pm 0.5$\tablefootmark{b} \\ \midrule
$\rm L_{1, CHeB}$ [$\mathrm{L_\odot}$]  &  65.5 & [65.3, 66] & [64.5, 99.2] & $37 \pm 4$\tablefootmark{c} \\ \midrule
$\rm T_{eff, 1, CHeB}$ [K]  &  5269 & [5013, 5966] & [4800, 26695] & $4710 \pm 50$\tablefootmark{c} \\ \midrule
\ce{^{12}C/^{13}C}  &  31 & [24, 35] & [13, 48] & $25 \pm 5$\tablefootmark{d} \\ \midrule
\ce{^{12}C/^{14}N}  &  1.4 & [1.1, 1.5] & [0.6, 2.0] & $1.4 \pm 0.2$\tablefootmark{d} \\ \midrule
\ce{^{14}N/^{16}O}  &  0.23 & [0.22, 0.26] & [0.19, 0.33] & $0.17 \pm 0.04$\tablefootmark{d} \\ \bottomrule
\end{tabular}
\tablebib{
\tablefoottext{a}{\citet[][]{Brewer2016};}
\tablefoottext{b}{\citet[][]{Handberg2017};}
\tablefoottext{b}{\citet[][]{Matteuzzi2023};}
\tablefoottext{d}{\citet[][]{Carlberg2015}}
}
\end{table}

\begin{figure*}
\centering
\includegraphics{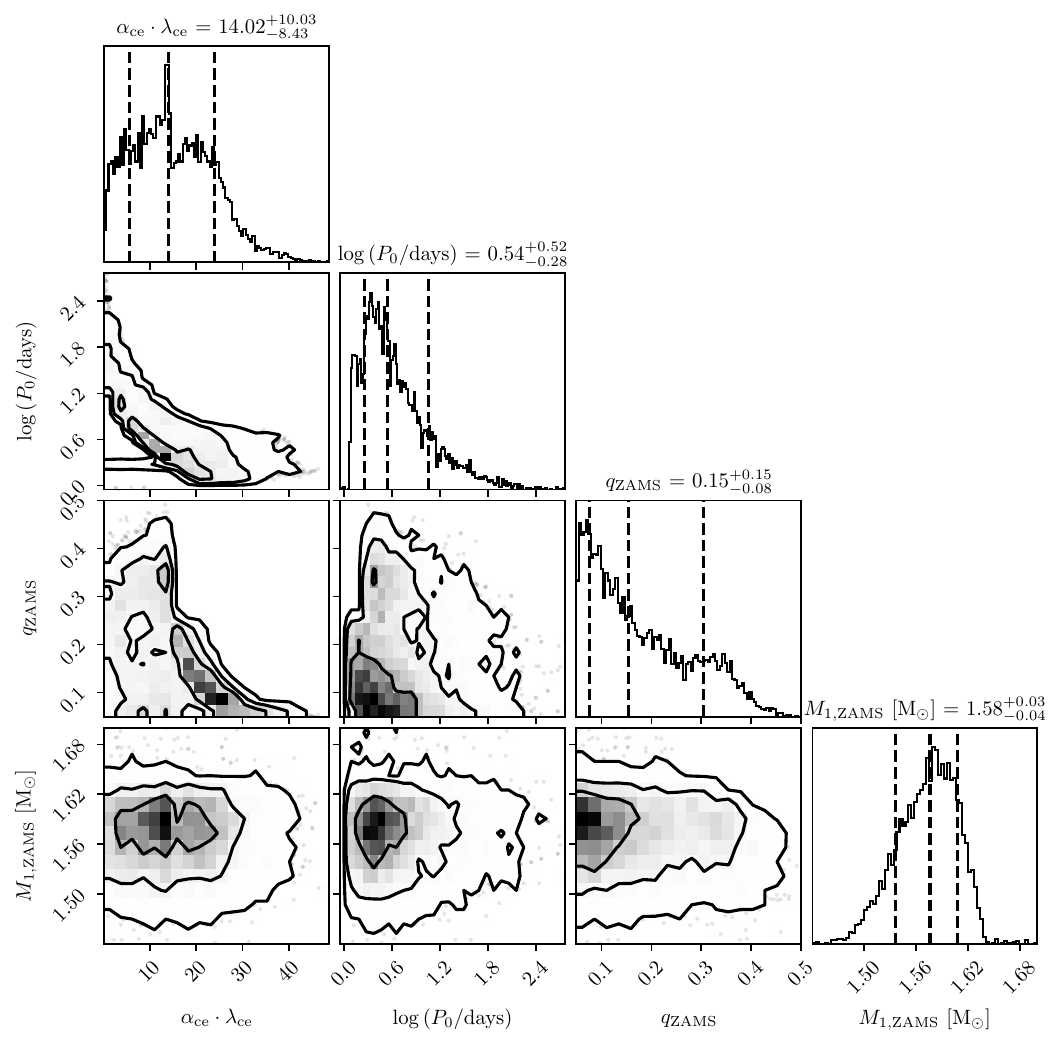} 
\caption{Corner plot showing the posterior density distributions of the subsample described in Section \ref{sec:results_subsample}. The contours are referred to 1, 2, and 3-$\sigma$, respectively.}
\label{fig:cornerplot_dynesty_remove_high_CN_L1}
\end{figure*}
In Section \ref{sec:chemistry} we explain that an interaction between a HeWD and a low-mass MS star can produce a low-mass CHeB star with the same mass and age as KIC4937011, but a higher \ce{^{12}C/^{14}N}. In this Section we want to exclude from the analysis all the MC simulations that predict stars much brighter and carbon-enriched than observations suggest, giving us more credible formation channels. Considering Figure \ref{fig:core_distrib1} and \ref{fig:chemistry}, we decide to exclude stars with luminosity and \ce{^{12}C/^{14}N} higher than $100 \, L_\odot$ and 2.5, respectively, because these thresholds are at least 5.5-$\sigma$ away from the observed values\footnote{The most credible formation channel remains the same as long as we choose a \ce{^{12}C/^{14}N} threshold that is still able to separate the two main chemical peaks. The luminosity threshold is necessary in order to exclude stars approaching the beginning of the EAGB.}. 13.67\% of our full sample of MC simulations (i.e. 8917 out of 65225 simulations) composes this subsample, thus, it is a non-negligible part of the full sample and we have sufficiently high number of MC simulations to calculate statistics\footnote{We check this by randomly sampling one third of our full sample. Even with a reduced number of MC simulations we infer the same results.}. In Fig. \ref{fig:cornerdyn_teff1_sub} the corner plot showing the age and primary mass density distributions for the subsample (see also Table \ref{tab:subsample}), compared with the observed values for KIC4937011 (red lines). We see that our modelling tends towards higher ages than the full sample, but within the errors the observed current age and mass of KIC4937011 are still correctly fitted.

\subsubsection{New posterior density distributions}
\label{sec:new_posteriors}
\begin{figure*}
\centering
\includegraphics{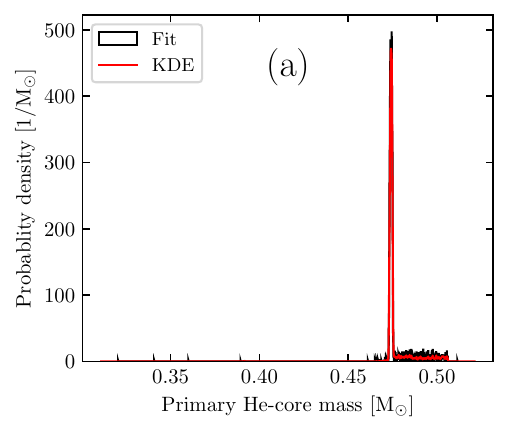} 
\includegraphics{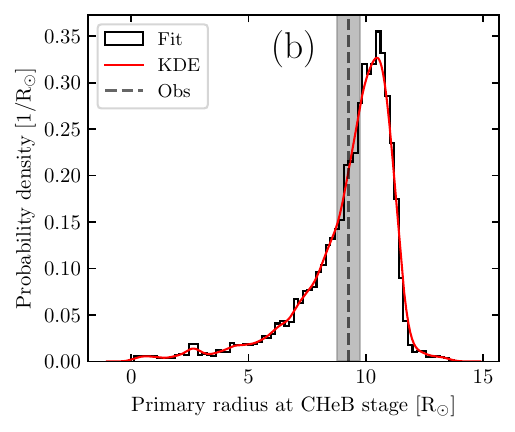}
\includegraphics{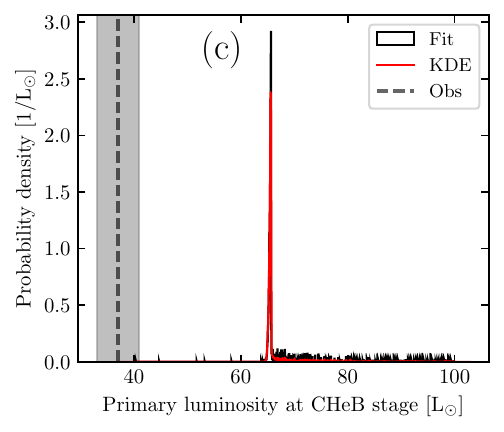}
\includegraphics{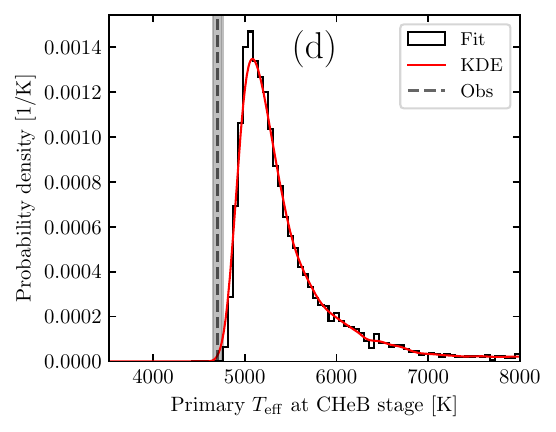}
\caption{Same as Figure \ref{fig:core_distrib1} but for the subsample described in Section \ref{sec:results_subsample} (see aso Table \ref{tab:subsample}). Also here effective temperatures above 8000 K have not been included in the figure for illustrative purposes only. As explained in the Appendix \ref{app:Likelihood}, we take as the reference time for the CHeB of each MC simulation the age that gives the highest likelihood.}
\label{fig:core_distrib1_sub}
\end{figure*}

\begin{figure}
\centering
\includegraphics{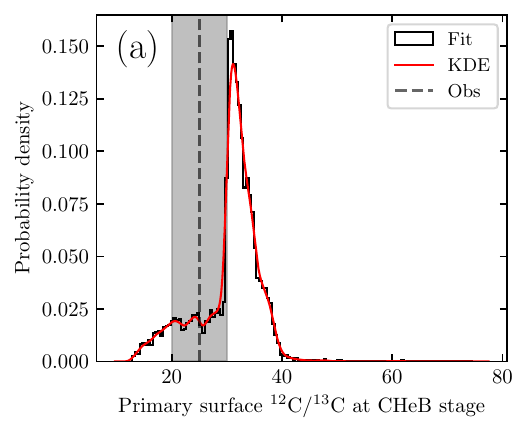} 
\includegraphics{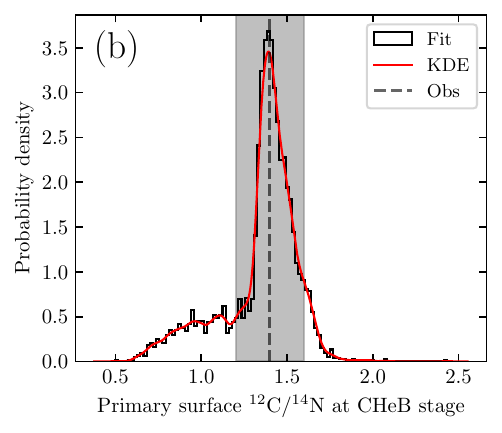} 
\includegraphics{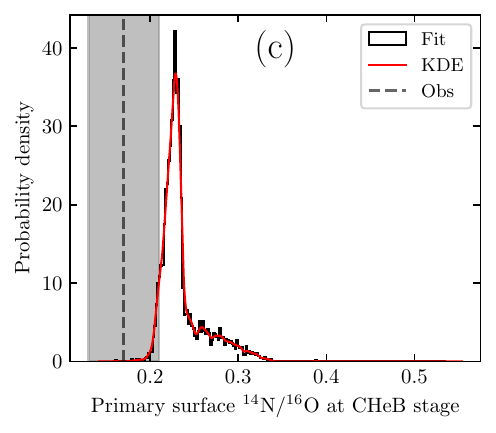} 
\caption{Primary star \ce{^{12}C/^{13}C}, \ce{^{12}C/^{14}N} and \ce{^{14}N/^{16}O} density distributions at the CHeB stage for the subsample described in Section \ref{sec:results_subsample}. In particular, we have the histogram (black lines) and the Kernel Density Estimation with a Gaussian kernel (red lines). The dashed line and the grey area represent the observed values of KIC4937011 and their 1-$\sigma$ errors (see also Table \ref{tab:fullsample}). As explained in the Appendix \ref{app:Likelihood}, we take as the reference time for the CHeB of each MC simulation the age that gives the highest likelihood.}
\label{fig:12C_over_13C_sub}
\end{figure}

In Fig.~\ref{fig:cornerplot_dynesty_remove_high_CN_L1}, \ref{fig:core_distrib1_sub} and \ref{fig:12C_over_13C_sub} the estimated posterior density distributions for the subsample (see also Table \ref{tab:subsample}). The anti-correlation between $\alpha_{\rm ce} \cdot \lambda_{\rm ce} $ and $\log \, P_0$ still holds, and both distributions have similar 99.7\% credible intervals and medians as the full sample (see Fig.~\ref{fig:cornerplot_dynesty_remove_high_CN_L1}, Tables \ref{tab:fullsample}, \ref{tab:subsample}). The same conclusion comes for the 99.7\% credible interval of $q_{\rm ZAMS}$, but not for its median value. Indeed, this density distribution in the subsample is skewed towards lower values of $q_{\rm ZAMS}$ compared to the full sample (see Tables \ref{tab:fullsample}, \ref{tab:subsample}). The density distribution of $M_{\rm 1, ZAMS}$ is also skewed towards lower values in the subsample than in the full sample, and it has narrower credible intervals than the full sample (see Tables \ref{tab:fullsample}, \ref{tab:subsample}). However, the $M_{\rm 1, ZAMS}$ distribution in the subsample is still consistent, within the errors, with the observed average mass of RGB stars in NGC6819, suggesting the presence of a fast evolution after the CE phase. We want to highlight that the $q_{\rm ZAMS}$ posterior distribution is also in the subsample very close to the lower limit we put in the prior, suggesting the same conclusions we drew in Section \ref{sec:results_fullsample}. Finally, as discussed in Section \ref{sec:results_fullsample}, the high $\alpha_{\rm ce} \cdot \lambda_{\rm ce} $ values observed in the subsample can be interpreted as an indication of recombination energy, dust formation, and nuclear burning influencing the CE ejection process.

The posterior density distributions of helium-core mass, radius, luminosity, and effective temperature are different in the subsample than in the full sample, they have narrower 99.7\% credible intervals, and they are more compatible with the observations (see Fig.~\ref{fig:core_distrib1_sub}, Tables \ref{tab:fullsample}, \ref{tab:subsample}). The helium-core mass distribution is consistent with the theoretical distribution expected for RC stars \citep[][]{Girardi2016}, and it is consistent with the asteroseismic observations of KIC4937011 \citep{Handberg2017}. The radius, luminosity, and effective temperature distributions are a separate case, because they tend to be systematically skewed towards higher values (see Section \ref{sec:core_distrib1}) than observations and modern evolutionary models (which are more compatible with observations, see Figure \ref{fig:HRD}). In fact, even if the median radius is consistent with the observations, within errors, the median luminosity and effective temperature are at least 7-$\sigma$ away from the observed values.
However, these systematic effects do not limit our inference in the formation channel, because they are not used in the likelihood (Appendix \ref{app:Likelihood}), thus, they are not used to discriminate between models. This is the reason why we decide not to rely on effective temperature, luminosity, and radius to perform a best fit to the observed values in KIC4937011.

Regarding the surface chemical composition distributions (Figure \ref{fig:12C_over_13C_sub} and Table \ref{tab:subsample}), the \ce{^{12}C/^{13}C} and \ce{^{12}C/^{14}N} posterior distributions are consistent, within the errors, with the observed values. The \ce{^{14}N/^{16}O} posterior distribution tends towards higher values than observations, but with a median of the distribution that is just about 1.2-$\sigma$ away from the observed value.

\subsubsection{KIC4937011's most credible formation channel}
\label{sec:formation_channel}
\begin{figure*}
\centering
\includegraphics[width=\textwidth]{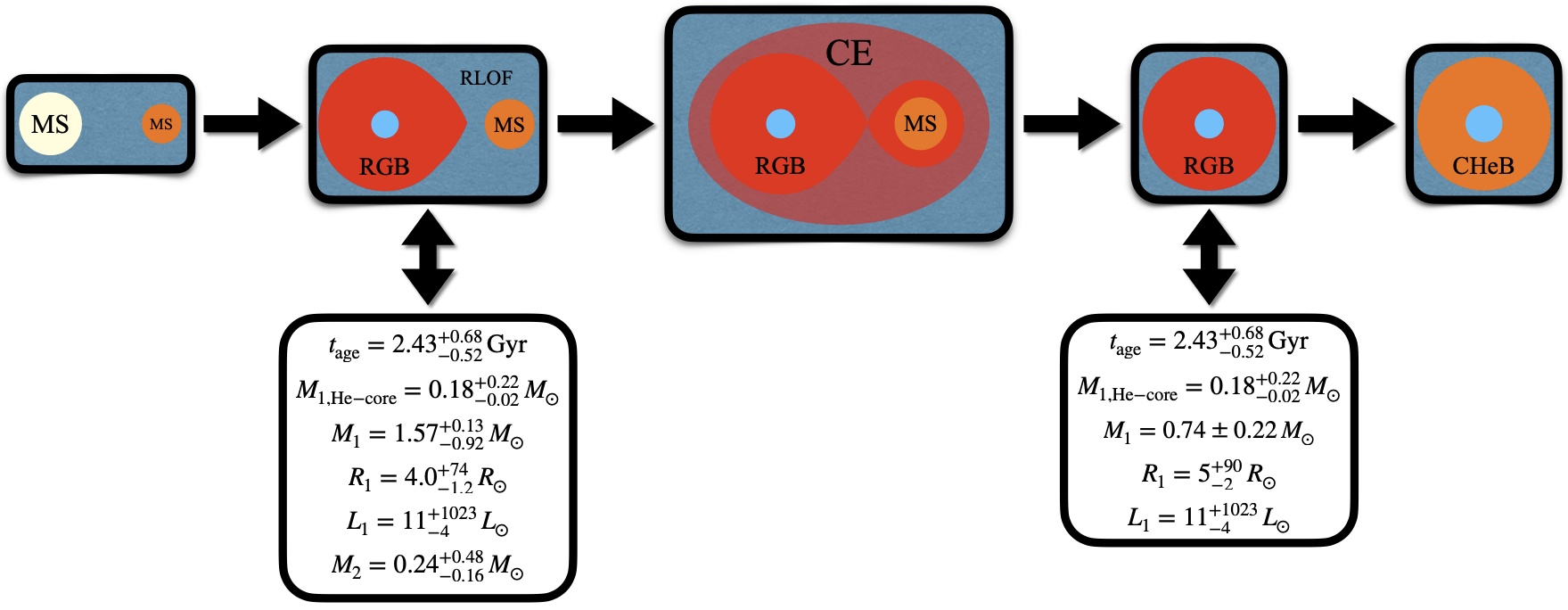} 
\caption{Cartoon showing the most credible formation scenario for KIC4937011. Medians and their 99.7\% credible intervals just before and after the CE are also shown.}
\label{fig:cartoon}
\end{figure*}
In Section \ref{sec:chemistry} we discuss the dichotomy visible in the \ce{^{12}C/^{14}N} distribution of the full sample and how the interaction between a HeWD star and a MS star is the main cause of the narrow peak at high \ce{^{12}C/^{14}N}. Even in the subsample (in which we are looking at the lowest values of the \ce{^{12}C/^{14}N} distribution) nearly all the MC simulations (99.92\% of the subsample, i.e. 8910 out of 8917 simulations) share a similar formation scenario (Fig. \ref{fig:cartoon}). As explained in Section \ref{sec:modeling} and visible in Fig. \ref{fig:cartoon}, we start from two ZAMS stars in a circular orbit. If we consider the 99.7\% credible intervals of Table \ref{tab:subsample}, we have a mass of the primary star between $1.46 \, M_\odot $ and $1.71 \, M_\odot $, a mass of the companion below $0.71$ \msol, and an orbital period between 1.15 days and 251 days. Such a close binary starts a RLOF when the primary goes in the RGB phase or in the HG phase. All these MC simulations predict an unstable RLOF and, thus, a CEE phase arises. Very differently to Section \ref{sec:chemistry} is the post-CEE phase result, because we have not a HeWD with a companion MS star. We have instead a RGB-like star with an evolved helium-core between $0.16$ \msol and $0.40$ \msol (99.7\% credible intervals, see Figure \ref{fig:cartoon}) and a smaller envelope than before the CEE phase (the 99.7\% credible interval for the mass of the star is between $0.52$ \msol and $0.96$ \msol after the CEE phase, see Figure \ref{fig:cartoon}). Indeed, we have a median ejection of $\approx 1.1$ \msol from the system and contemporary the merger of the companion with the helium-core of the primary star. During the CEE phase we have dynamical mixing effects owing to the spiral-in process that completely mix the envelope with material coming from the evolved primary star. After the CEE phase the RGB-like star has an envelope with the low \ce{^{12}C/^{14}N} value we observe today. Finally, this star reaches the CHeB stage after helium flashes as if it were a single star of mass $\approx 0.71$ \msol, consistently with observations.
Therefore, this formation channel "naturally" predicts the low \ce{^{12}C/^{14}N} value compatible with RC star of $\approx 1.6$ \msol (Section \ref{sec:obs}). Moreover, the $R_{\rm 1,CHeB}$, $L_{\rm 1,CHeB}$, and $\rm T_{eff}$ posterior distributions we found after all these binary interactions are compatible with a $\approx 0.7$ \msol CHeB star that evolves in isolation, indicating that our findings are self-consistent within the {\scshape binary\_c v2.2.3} framework.

However, with this formation channel is difficult to explain the high lithium abundance we observe today in KIC4937011. Insights into this conundrum may be gleaned from studies of red novae, which are thought to be the direct product of a CEE phase with a merger \citep[][]{Tylenda2006,Pastorello2019}. Some of these red novae are lithium-rich \citep[][]{Kaminski2023}, suggesting the involvement of mechanisms capable of synthesizing and mixing lithium during the CEE phase.

Between the post-CEE phase and the CHeB stage the star evolves in isolation for nearly 150 Myr (maximum a posteriori probability estimate, corresponding to stars with a helium-core mass of about $0.18$ \msol), which is very short compared to the cluster age. We check whether this time and this formation channel are compatible with more state-of-the-art evolutionary codes of single stellar evolution. Using the {\scshape MESA v11701} \citep[Modules for Experiments in Stellar Astrophysics;][]{Paxton2011,Paxton2019} tool we test the pre-CEE and post-CEE phase conditions.
We adopt as a reference solar mixture that from \citet{Asplund2009}, and high- and low-temperature radiative opacity tables were computed for the solar specific metal mixture. Envelope convection is described by the mixing length theory \citet{Cox1968}; the corresponding $\alpha_\mathrm{MLT}$ parameter, the same for all the models, is derived from the solar calibration with the same physics. Below the convective envelope, we add a diffusive undershooting \citep{Herwig2000} with a size parameter $f = 0.02$ \citep[see][]{Khan2018}. Extra mixing over the convective core limit during the CHeB phase is treated following the formalism by \citet{Bossini2017}.
For the pre-CEE phase we adopt a $1.60$ \msol star with $Z = 0.022$ and $Y = 0.28$ until it achieves a $0.18$ \msol helium-core mass in the RGB phase, which takes nearly 2.33 Gyr. For the post-CEE phase we use models with masses between $0.65$ \msol and $0.80$ \msol (i.e. a mass range compatible with the 68\% credible interval of $M_{\rm 1, CHeB}$ in Table \ref{tab:subsample}), and $Z = 0.022$, $Y = 0.28$. We do not include any mass loss in these MESA models. It takes about 500 Myr for a $0.65$ \msol star to go from a RGB phase with a helium-core mass of $0.18\, M_\odot$ to the CHeB stage, and about 370 Myr for a $0.80$ \msol star. This means that in MESA models this formation channel would take at least 2.70 - 2.83 Gyr, which is about 1.6-$\sigma$ away from the observed value. However, if we consider not just a single value of the helium-core mass, but a distribution of values just before the CEE phase, it is also possible in MESA to have a final age consistent with {\scshape binary\_c v2.2.3}. For example, it takes 2.48 Gyr for a $1.60$ \msol star in MESA to achieve a helium-core mass of $0.22$ \msol, and other about 150 Myr for a star between $0.65$ \msol and $0.80$ \msol to ignite helium in the core.

\begin{figure}
\centering
\includegraphics{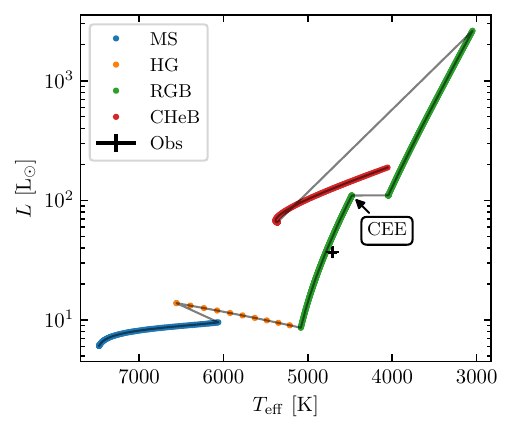} 
\caption{HRD of a primary star in the subsample from the ZAMS to the end of the CHeB stage done with {\scshape binary\_c v2.2.3}. The CEE begins when the primary star has a helium-core mass of about $0.279$ \msol (arrow in the figure), and it ends with the ejection of $1.12$ \msol of material from the system. The simulation begins with a $1.60$ \msol ZAMS primary star orbiting a $0.25$ \msol companion in almost 32 days. The primary star begins the CHeB stage at the age of 2.41 Gyr and a mass of $0.70$ \msol. In black KIC4937011's observed values (see also Table \ref{tab:subsample}).}
\label{fig:HRD_binary_c}
\end{figure}
An example of {\scshape binary\_c v2.2.3} HRD of a primary star with such a formation channel is in Figure \ref{fig:HRD_binary_c}. This specific simulation begins with a $1.60$ \msol ZAMS primary star orbiting a $0.25$ \msol companion in almost 32 days. The CEE begins when the primary star has a helium-core mass of about $0.279$ \msol, and it ends with the ejection of $1.12$ \msol of material from the system. The post-CEE star begins the CHeB stage at the age of 2.41 Gyr and a mass of $0.70$ \msol.

\section{Discussion and conclusions}
\label{sec:conc}
In this work we focus on the single metal-rich ($Z \approx \Zsol$), Li-rich, low-mass, CHeB star KIC4937011, which is a member of the star cluster NGC 6819 (turn-off mass of $\approx 1.6$ \msol, i.e. age of $\approx 2.4$ Gyr). This star has $\approx 1$ \msol less mass than expected for its age and metallicity, thus, it could be the result of a binary interaction or of the poorly understood mass loss mechanism along the red-giant branch. To infer formation scenarios, we adopt a Bayesian approach using the {\scshape binary\_c v2.2.3} code coupled with the Dynamic Nested Sampling approach contained in the {\scshape dynesty v2.1.1} package.
The final conclusions are summarised here:
\begin{enumerate}
\item This star is the result of a common-envelope evolution phase in which the companion does not survive. All the MC simulations considered have one common envelope phase within the final stage of the CHeB phase of the primary star, and only a negligible fraction of simulations (0.006\%) experiences two common envelope phases.

\item Considering a subsample composed by CHeB primary stars with luminosity below $100 \, L_\odot$ and \ce{^{12}C/^{14}N} below 2.5, we find that the highest peak in their helium-core mass density distribution is at $0.4743$ \msol. This is in line with the expectations for RC stars, and with the observations of KIC4937011. The \ce{^{12}C/^{13}C} and \ce{^{12}C/^{14}N} distributions are consistent, within the errors, with the observations. The \ce{^{14}N/^{16}O} posterior distribution tends towards higher values than observations, but with a median of the distribution that is just about 1.2-$\sigma$ away from the observed value. The effective temperatures, luminosities, and radii distributions are systematically higher than observations. This is expected, because for such low-mass stars ($\approx 0.7$ \msol) the \citet{Pols1998} evolutionary tracks used in {\scshape binary\_c v2.2.3} deviates significantly from more modern evolutionary tracks \citep[][]{Girardi2016}. However, these systematic effects do not limit our inference in the formation channel, because they are not used in the likelihood, thus, they are not used to discriminate between models.

\item The most credible formation channel is summarised in Figure \ref{fig:cartoon}. We start from two ZAMS stars in a circular orbit. In all the MC simulations considered, a RLOF begins when the companion is still in the MS phase, but with a primary star that goes in the RGB phase or in the HG phase. This RLOF is dynamically unstable and, thus, a CEE phase arises. During this CEE phase we have dynamical effects that induce a mixing of the envelope of the primary star with the MS companion, forming the chemical pattern we observe today. The final effect of this CEE phase is a median ejection of $\approx 1.1$ \msol of material from the system and contemporary the merger of the companion with the helium-core of the primary star. The post-CEE phase product is an RGB-like star of $\approx 0.71$ \msol with an evolved He core. Finally, the star reaches the CHeB stage after helium flashes as if it were a single star. This formation channel is consistent with MESA models of single star evolution before and after the CEE phase if we take systematic effects into account.

\end{enumerate}
It is worth mentioning that the same formation channel could form sdB star or metal-rich RR Lyrae \citep{Bobrick2024}, because a non-negligible fraction of low-mass CHeB stars has effective temperatures reaching about 27000 K. This means that this channel potentially provides another piece of the puzzle in the sequence between RC and sdB stars, or other stripped stars. 

In future we will analyse other very-low mass CHeB stars in open clusters with the same technique, because a better statistics of such objects is necessary to verify the formation channel as an universal feature. Furthermore, synthetic populations of \kepler stars provide valuable insights into the overall probability of such an evolutionary scenario (Mazzi et al., in prep.). We will also improve the posterior distributions by using detailed evolutionary models within the likelihood in order to compute $\lambda_{\rm ce}$ and reduce systematic effects. An analysis of the activity-sensitive He I 10830 Å absorption triplet would also be interesting in order to better study the current mass loss in KIC4937011 and in order to put this star in the wider context of the Li-rich RC stars \citep[][]{Sneden2022}.

\begin{acknowledgements}
We are grateful to Raffaele Pascale and Payel Das for useful discussions. We are also grateful to Rasmus Handberg for sharing the data used in \citet{Handberg2017} with us. We thank the anonymous referee for their helpful comments that improved the quality of the manuscript. This work has made use of data from the European Space Agency (ESA) mission Gaia (\url{https://www.cosmos.esa.int/gaia}). MM, AM, KB, JM, MT acknowledge support from the ERC Consolidator Grant funding scheme (project ASTEROCHRONOMETRY, \url{https://www.asterochronometry.eu}, G.A. n. 772293). We made use of the Python code {\scshape corner} \citep{Foreman-MackeyCorner}, and of the Python code code {\scshape KDEpy} (\url{https://zenodo.org/doi/10.5281/zenodo.2392267}).
\end{acknowledgements}

\section*{Data Availability}
The data underlying this article will be shared on reasonable request to the corresponding author.



\bibliographystyle{aa}
\bibliography{paperce} 




\begin{appendix}

\section{Likelihood and prior functions}
\label{app:Likelihood}
As discussed in Section \ref{sec:modeling}, we calculate the likelihood function given the current evolutionary phase (CHeB stage), mass ($M_{\rm 1, \, obs} \pm \sigma_{\rm 1, \, obs} = 0.71 \pm 0.08$ \msol) and age ($t_{\rm age, \, obs} \pm \sigma_{\rm age, \, obs} = 2.38 \pm 0.27$ Gyr) of KIC4937011. We employ the multivariate normal likelihood
\begin{equation}
    \mathcal{N}({\bf x} | \boldsymbol{\mu},\boldsymbol{C}) = \max_{ \boldsymbol{t_{\rm age, \, mod}}} \frac{1}{2 \pi \sqrt{| {\bf C}|} } \exp{ \left[ - \frac{1}{2} \left( {\bf x} -  \boldsymbol{\mu} \right)^T \boldsymbol{C}^{-1} \left( {\bf x} -  \boldsymbol{\mu} \right) \right] }
    \label{eq:likelihood}
\end{equation}
with ${\bf x} = [t_{\rm age, \, obs}; M_{\rm 1, \, obs}]$, $\boldsymbol{\mu} = [\boldsymbol{t_{\rm age, \, mod}}; \boldsymbol{M_{\rm 1, \, mod}}]$, and $\boldsymbol{C} = \diag{\sigma_{\rm age, \, obs}^2;  \sigma_{\rm 1, \, obs}^2}$. It is not possible to univocally associate an age with the CHeB stage, because several ages ($\boldsymbol{t_{\rm age, \, mod}}$ and the corresponding mass $ \boldsymbol{M_{\rm 1, \, mod}}$ are not scalars) correspond to the same evolutionary phase. Therefore, for each evolutionary track we calculate the likelihood corresponding to all ages in the same evolutionary phase and we keep the age that leads to the highest likelihood as a characteristic age to associate at the CHeB stage of that evolutionary track.

In Table \ref{tab:priorsrange} the intervals of the priors. From a comparison with Table \ref{tab:fullsample} and \ref{tab:subsample} we can note that three free parameters are well constrained within the intervals. This suggests that it is not necessary to widen these ranges because we explore the free-parameters space thoroughly. However, $q_{\rm ZAMS}$ has a lower bound of the posterior distribution that touches our chosen prior (see discussion in Section \ref{sec:results_fullsample} and \ref{sec:results_subsample}).
\begin{table}
\centering
\caption{Intervals of the priors used in Section \ref{sec:modeling}. A comparison with Table \ref{tab:fullsample} and \ref{tab:subsample} shows that only $q_{\rm ZAMS}$ has a lower bound of the posterior distribution that touches the prior.}
\label{tab:priorsrange}
\begin{tabular}{@{}lll@{}}
\toprule
Parameter & prior &  interval\\ \midrule
$\alpha_{\rm ce} \cdot \lambda_{\rm ce}$ & Uniform & $[0.0, 100.0]$\\ \midrule
$\log \left(P_0 / \mathrm{days}\right)$ & Uniform& $[-10.0, 10.0]$ \\ \midrule
$q_{\rm ZAMS}$  & Uniform& $[0.08 \, \mathrm{M_\odot}/M_{\rm 1, ZAMS}, 1.0]$ \\ \midrule
$M_{\rm 1, ZAMS}$ [$\mathrm{M_\odot}$]  & \citet{Chabrier2003} & $[0.08, \infty)$ \\ \bottomrule
\end{tabular}
\end{table}

We also explore the posterior distributions using the same priors and likelihood, but with the Markov Chain Monte Carlo algorithm (MCMC) present in the {\scshape emcee} Python package \citep{Foreman-Mackey2013}. We obtain results consistent with Section \ref{sec:results}.


\end{appendix}

\end{document}